\RequirePackage{amsmath}
\documentclass[runningheads,a4paper]{llncs}

\usepackage[utf8]{inputenc} 
\usepackage{xcolor}
\usepackage[color=blue!20, linecolor=blue!80!black]{todonotes}
\usepackage[english]{babel}
\usepackage[utf8]{inputenc}
\DeclareUnicodeCharacter{00A0}{ }
\usepackage{hyperref}
\usepackage{xspace}
\usepackage{amsmath} 
\usepackage[ruled,vlined]{algorithm2e}
\usepackage[capitalise,nameinlink]{cleveref}
\usepackage[binary,squaren]{SIunits} 
\usepackage{amssymb}
\usepackage{listings}
\usepackage{dsfont}
\usepackage{multicol}
\usepackage{cite}
\usepackage{float}
\usepackage{booktabs}
\usepackage{cleveref}
\usepackage{pgfplots}
\usepackage{caption}
\usepackage{subcaption}
\usepackage{soul}
\usepackage{mathtools}
\usepackage{xparse}
\usepackage[inline]{enumitem}
\usepackage{graphicx}

\usepackage{booktabs} 
\usepackage{xspace} 
\usepackage{lstautogobble}  

\usepackage{underscore}

\usepackage{pgfplots}
\usepgfplotslibrary{colorbrewer}
\pgfplotsset{compat = 1.15, cycle list/Set1-8} 
\usetikzlibrary{pgfplots.statistics, pgfplots.colorbrewer} 
\usepackage{pgfplotstable}
\usepackage{filecontents}

\usetikzlibrary{shapes, arrows, positioning, calc}
\makeatletter
\newcommand*{\etal}{%
    \@ifnextchar{.}%
        {et~al}%
        {et~al.\@\xspace}%
}
\newcommand*{\ie}{i.e.\@\xspace}
\newcommand*{\eg}{e.g.\@\xspace}

\makeatother

\newcommand\donotshow[1]{}

\newcommand*{\set}[1]{\{{#1}\}}

\NewDocumentCommand{\twopartdef}{ m m m o}{
  \left\{
    \begin{array}{ll}
      #1 & \mbox{if } #2 \\
      #3 & \IfNoValueTF{#4}{\text{otherwise}}{\mbox{if } #4}
    \end{array}
  \right.
}
\NewDocumentCommand{\threepartdef}{m m m m m o}{
  \left\{
    \begin{array}{lll}
      #1 & \mbox{if } #2 \\
      #3 & \mbox{if } #4 \\
      #5 & \IfNoValueTF{#6}{\text{otherwise}}{\mbox{if } #6}
    \end{array}
  \right.
}
\NewDocumentCommand{\longthreepartdef}{m m m m m m m o}{
  \left\{
    \begin{array}{lll}
      #1 & \mbox{if } #2 \\
         & #3\\
      #4 & \mbox{if } #5 \\
         & #6\\
      #7 & \IfNoValueTF{#8}{\text{otherwise}}{\mbox{if } #8}
    \end{array}
  \right.
}
\NewDocumentCommand{\fourpartdef}{m m m m m m m o}{
  \left\{
    \begin{array}{llll}
      #1 & \mbox{if } #2 \\
      #3 & \mbox{if } #4 \\
      #5 & \mbox{if } #6 \\
      #7 & \IfNoValueTF{#8}{\text{otherwise}}{\mbox{if } #8}
    \end{array}
  \right.
}

\newcommand*\rtlola{\textsc{RTLola}\xspace}

\newcommand*\vhdl{\text{VHDL}\xspace}

\newcommand*{\outputmark}[1]{{\ensuremath{#1^{\scriptscriptstyle\uparrow}}}}

\newcommand*{\generalmark}[1]{{\ensuremath{#1^{\scriptscriptstyle\mathord{-}}}}}

\newcommand*{\streamout}[1]{{\outputmark{s_{#1}}}\xspace}
\newcommand*{\streamany}[1]{{\generalmark{s_{#1}}}\xspace}

\newcommand*{\outputpt}[1]{\ensuremath{\streamout{#1}.\mathit{pt}}}

\newcommand*\evalorder{\mathrel{\prec}}

\definecolor{CommentColor}{RGB}{42,0.0,255} 

\definecolor{bluekeywords}{rgb}{0.13, 0.13, 1}
\definecolor{greentypes}{rgb}{0, 0.5, 0}
\definecolor{redstrings}{RGB}{171, 114, 2}
\definecolor{graynumbers}{rgb}{0.5, 0.5, 0.5}
\definecolor{goldcomments}{rgb}{0.6, 0.4, 0.08}
\definecolor{monitorblue}{RGB}{18, 163, 38}

\lstdefinelanguage{Lola}{
  keywords=[0]{input, output, trigger, import},
  keywordstyle=[0]\bfseries\color{bluekeywords},
  keywords=[1]{if, then, else, aggregate, defaults, offset, filter, hold},
  keywords=[2]{Int8, Int16, Int32, Int64, UInt8, UInt16, UInt32, UInt64, Bool, Float16, Float32, Float64, @1Hz, @2Hz, @5Hz, @10Hz, @100mHz, @1kHz},
  keywordstyle=[2]\color{greentypes},
    sensitive=false,
    comment=[l]{//},
    morecomment=[s]{/*}{*/},
    morestring=[b]',
    morestring=[b]"
}

\colorlet{eventcolor}{green!50!black}
\colorlet{periodiccolor}{blue!50!black}
\tikzstyle{event} = [draw=eventcolor, thin, fill opacity=.3, pattern=north west lines, pattern color=eventcolor]
\tikzstyle{periodic} = [draw=periodiccolor, thin, fill opacity=.3, pattern=north east lines, pattern color=periodiccolor]
\tikzstyle{signalname} = [text width=10em, minimum height=2em]
\tikzstyle{nameright} = [signalname, align=left]
\tikzstyle{namecenter} = [minimum height=2em, align=center, rotate=65]
\tikzstyle{nameleft} = [signalname, align=right]

\begin{document}
\hyphenation{Time-stamp}
\hyphenation{Au-to-no-mous}

\title{Automatic Optimizations for \\ Stream-based Monitoring Languages\thanks{This work was partially supported by the German Research Foundation (DFG) as part of the Collaborative Research Center Foundations of Perspicuous Software Systems (TRR 248, 389792660), and by the European Research Council (ERC) Grant OSARES (No. 683300).}}

\author{Jan Baumeister\inst{1}\orcidID{0000-0002-8891-7483} \and
Bernd Finkbeiner\inst{1}\orcidID{0000-0002-4280-8441} \and
Matthis Kruse\inst{2}\orcidID{0000-0003-4062-9666}\and \\
Maximilian Schwenger\inst{1}\orcidID{0000-0002-2091-7575}
}
\authorrunning{Baumeister et al.}
\titlerunning{Automatic Optimizations for Stream-based Specification Languages}
\institute{%
  CISPA Helmholtz Center for Information Security,
  Saarland Informatics Campus\\
  66123 Saarbr\"ucken, Germany \\
  \email{\{jan.baumeister, finkbeiner, maximilian.schwenger\}@cispa.saarland}
  \and
  Saarland University, Saarland Informatics Campus\\
  66123 Saarbr\"ucken, Germany \\
  \email{matthis.kruse@cs.uni-saarland.de} 
}
\maketitle

\begin{abstract}
Runtime monitors that are specified in a stream-based monitoring language tend to be easier to understand, maintain, and reuse than those written in a standard programming language.
Because of their formal semantics, such specification languages are also a natural choice for safety-critical applications.
Unlike for standard programming languages, there is, however, so far very little support for automatic code optimization.
In this paper, we present the first collection of code transformations for the stream-based monitoring language \rtlola.
We show that classic compiler optimizations, such as Sparse Conditional Constant Propagation and Common Subexpression Elimination, can be adapted to monitoring specifications.
We also develop new transformations --- Pacing Type Refinement and Filter Refinement --- which exploit the specific modular structure of \rtlola as well as the implementation freedom afforded by a declarative specification language.
We demonstrate the significant impact of the code transformations on benchmarks from the monitoring of unmanned aircraft systems (UAS).
\end{abstract}

\section{Introduction}

The spectrum of languages for the develpment of monitors ranges from
standard programming languages, like Java and C\nolinebreak[4]\hspace{-.05em}\raisebox{.4ex}{\tiny\bf ++}, to formal logics
like LTL and its many variations. The advantage of programming
languages is the universal expressiveness and the availability of
modern compiler technology; programming languages lack, however, the
precise semantics and compile-time guarantees needed for
safety-critical applications. Formal logics, on the other hand, are
sufficiently precise, but have limited expressiveness.  A good trade-off  
between the two extremes is provided by stream-based monitoring languages like
\rtlola. Stream-based languages have the expressiveness of a programming
language, and, at the same time, the formal semantics and compile-time
guarantees of a formal specification language.

For standard programming languages, the development of effective code
optimizations is one of the most fundamental research questions.  By
contrast, there is, so far, very little support for the automatic
optimization of monitoring specifications.  In this paper, we present
the first collection of code transformations for the stream-based
monitoring language \rtlola~\cite{rtlolacavtoolpaper}.

Our starting point are compiler
optimizations known from imperative programming languages like
\emph{Sparse Conditional Constant Propagation} and \emph{Common
  Subexpression Elimination}.  Adapted to stream-based specifications,
such transformations allow the user to write code that is easy to read
and maintain, without the performance penalty resulting, for example,
from unnecessarily recomputing the value of subexpressions.

We also develop optimizations that are specific to stream-based monitoring.
Stream-based languages have several features that
make them a particularly promising target for code
optimization.
Stream-based languages are \emph{declarative} in the sense that it is
only the correct computation of the trigger conditions that matters
for the soundness of the monitor, not the specific order in which
intermediate data is produced. This leaves much more freedom for code
transformation than in an imperative language.  Another feature of
stream-based languages that is beneficial for code transformation is
that the write-access to memory is inherently \emph{local}: the
computation of a stream only writes once in its local memory while
potentially reading multiple times from other streams\footnote{This is
  related to the functional programming paradigm where function calls
  are \emph{pure}, \ie, free of side effects.}.  This means that
expressions used for the computation of one stream can be modified
without affecting the other streams.  Finally, our code optimization
exploits the clear \emph{dependency structure} of stream-based
specifications, which allows us to efficiently propagate type
changes made in one stream to all affected streams in the remainder of
the specification.
We present two transformations that specifically exploit these
advantages. \emph{Pacing Type Refinement} optimizes the points in time
when a stream value is calculated, eliminating the computation of
stream values that are never used. \emph{Filter Refinement} avoids
the unncecessary computation of expressions that appear in the scope of an
\emph{if} statement, ensuring that the expression is only evaluated
if the condition is actually true.

\rtlola specifications are used both in interpretation-based monitors~\cite{rtlolacavtoolpaper} and as the source language for compilers, for example to \vhdl~\cite{fpgalola}.
Our code transformations are applicable in both approaches, because the
transformations are applied already on the level of intermediate
representations (AST, IR).  In transpilation backends, the optimized
code is compiled one more time, and thus additionally benefits from the
standard compiler optimizations for the target platform.

A prime application area for our optimizations is the monitoring of unmanned aircraft systems (UAS)~\cite{rtlolacavindustrial}.
Monitoring aircraft involves complex computations, such as the crossvalidation of different sensor modules. The performance of the monitor implementation is critical,
because the on-board monitor is executed on a platform with limited computing power.
Our experience with the code transformations (for details see Section~\ref{sec:evaluation}) is very encouraging.

\subsection{Related Work}

This paper presents the first collection of code transformations for the stream-based monitoring language \rtlola.  There is, of course, a vast literature on
compiler optimization. For an introduction, we refer the reader to the
standard textbooks on compiler design and implementation
(cf.~\cite{10.5555/286076,dragonBook,Seidl:2012:OptEn}). Kildall~\cite{kildall}
gives a comprehensive overview on the classic code
transformations. The foundation for the code transformations is
provided by methods from program analysis such as abstract
interpretation~\cite{abstract-interpretation}.

The programming paradigm that most closely resembles stream-based monitoring languages like \rtlola is 
\emph{synchronous programming}. Examples of synchronous programming languages are  \textsc{Lustre}~\cite{halbwachs91synchronous}, \textsc{Esterel}~\cite{berry00foundations}, and \textsc{Signal}~\cite{gautier87signal}. These languages are supported by optimization techniques like the annotation-based memory optimization of \textsc{Lustre}~\cite{annotations-regalloc} and the low-level elimination of redundant gates and latches in \textsc{Esterel}~\cite{redundancy-elim}.
There are, however, important differences to the transformations presented in this paper. Our transformations work on the level of intermediate representations, which makes them  uniformly applicable to interpretation and compilation. The new \emph{Pacing Type} and \emph{Filter Refinements} furthermore exploit the specific modular structure of \rtlola as well as the much greater implementation freedom afforded by a declarative specification language. 

Our focus on \rtlola is motivated by recent work on \rtlola-based monitoring for UAS~\cite{rtlolacavindustrial} and other cyber-physical systems~\cite{fpgalola,rtlolacavtoolpaper}. It should be possible, however, to develop similar optimizations for other stream-based monitoring languages like  
TeSSLa\cite{tessla} and Striver\cite{striver}.

\section{RTLola}\label{sec:rtlola}
\rtlola\cite{rtlolaarxiv,rtlolacavtoolpaper} is a runtime monitoring framework.
In its core, it takes a specification in the eponymous specification language and analyzes whether and when input data violates the specification.
To this end, it interprets sequences of incoming data points as input streams.
The \rtlola stream engine then transforms these values according to stream expressions in the specification to obtain output streams.
The specification also contains trigger conditions, \ie, boolean expressions indicating whether a certain property is violated or not.
Stream expressions and trigger conditions depend either on input or output stream values.

Consider the following \rtlola specification.
\begin{lstlisting}
input gps: (Float64, Float64)
output gps_readings: Bool@1Hz := gps.aggregate(over:2s,using:count)
trigger gps_glitch < 10 "GPS sensor frequency < 5Hz"
\end{lstlisting}
The specification first declares an input stream with the name \lstinline{gps}.
The output stream \lstinline{gps_readings} analyzes the input stream by counting how many readings the monitor received within the last 2\second. 
This computation is a sliding window, so when the \lstinline{gps_readings} stream computes a new value at point in time $t$, \rtlola takes all data points of the \lstinline{gps} stream into account, which were received in the interval $[t-2\second, t]$.
The trigger then checks whether the number of GPS readings in such a $2\second$ interval falls below 10. 
If so, it raises an alarm such that the observed system can react accordingly \eg by initiating mitigation procedures.

\subsection{Type System}

Types in \rtlola are two-dimensional consisting of the value type and the pacing type.
The former is drawn from a set of types representable with a static amount of bits.  
The pacing type consists of two components:  an evaluation trigger and a filter condition.
The monitor will compute a new value for a stream as soon as the evaluation trigger occurred \emph{unless} the filter condition is false.
Let us ignore filter conditions for now.
The evaluation trigger can be a real-time frequency as was the case for \lstinline{gps_readings}.  
In this case, the stream is a \emph{periodic} stream.
Otherwise, the evaluation trigger is a positive boolean formula $\varphi$ over the set of input streams, in which case the stream is \emph{event-based}.
The reason behind this lies within the input model of \rtlola.
\rtlola assumes input values to arrive asynchronously, \ie, if a specification declares several input streams $\mathcal{I}$, an incoming data point $\mathcal{I}'$ can cover an arbitrary non-empty subset $\emptyset \neq \mathcal{I}' \subseteq \mathcal{I}$.
Only streams in $\mathcal{I}'$ receive a new value.
Thus, the monitor evaluates event-based streams with evaluation trigger $\iota$ iff $\mathcal{I}' \implies \iota$. 
I.e., it replaces all occurrence of the input stream name $i$ in $\iota$ by true if $\iota \in \mathcal{I}'$ and false otherwise.
Consequently, any input stream $i$ has evaluation trigger $\set{i}$ intuitively meaning ``$i$ will be extended when the system provides a new value for it.''
For event-based streams, the evaluation trigger is called the \emph{activation condition}.

Note that the type annotation of \lstinline{gps} in the previous example does not contain information about the pacing type at all.
In many cases, \rtlola infers the types of streams automatically based on the stream expression rendering type annotations largely optional.
While the type inference for value types is straight-forward because \rtlola requires input streams to have type annotations, the inference for pacing types is mainly based on stream accesses.
There are three kinds of stream accesses: synchronous, asynchronous, and aggregations.
If a stream $x$ accesses a stream~$y$ synchronously, then the evaluation of $x$ demands the $n$th-to-latest value of $y$ where $n$ is the \emph{offset} of the access.
This ties the evaluation of both streams together, so if $y$ has an evaluation frequency of $5\hertz$, $x$ cannot be evaluated more frequently, nor can $x$ be event-based.
Asynchronous accesses refer to the last value of a stream, no matter how old it may be.  
Here, the pacing of $x$ and $y$ remain decoupled.
Aggregating accesses --- such as the one in \lstinline{gps_readings} --- decouple the pacing as well.

Lastly, filter conditions are regular \rtlola expressions. 
Assume stream $x$ has the evaluation condition $\pi$ with filter $\phi$. 
Whenever $\pi$ is true, the monitor evaluates the filter $\phi$.
Only if the filter is true as well, the monitor evaluates the stream expression and extends $x$.

\subsection{Evaluation}

An \rtlola specification consists of input streams, output streams, and triggers.
The monitor for a specification computes a static schedule containing information on which a periodic stream needs to be computed at which point in time.
When such a point in time is reached or the monitor receives new input values, it starts an evaluation cycle.
Here, the monitor first determines which streams could be affected by checking their frequencies or activation conditions.
It then orders them according to an \emph{evaluation order} $\evalorder$.
Following this order, the monitor checks the filter condition of each stream.
If it evaluates to true, the monitor extends the stream by evaluating the stream expression to obtain a new value.

This process only works correctly if the evaluation order complies with the \emph{dependency graph} of the specification.
The annotated dependency graph is a directed multigraph consisting of one node for each trigger, stream, and filter condition.
Each edge in the graph represents a stream access in the specification.
For the evaluation order, only synchronous lookups matter:  if node $s$ access node $s'$ synchronously, $s'$ needs to be evaluated before $s$.

After the evaluation, the monitor checks whether a trigger conditions was true.  
If so, passes the information on to the system under scrutiny.
This constitutes the \emph{observable behavior} of the monitor, any other computation is considered internal behavior.
Consequently, any computation that does not impact a trigger condition is completely irrelevant.

This is just a rough outline of \rtlola.  
For more information refer to \cite{maxmaster}.

\begin{remark}[Transformations Preserve Observable Behavior]
The point behind the compiler transformations presented in this paper is to improve the running time and thus decrease the latency between the occurrence and report of a violation.
Yet, the correctness, \ie, the observable behavior of the monitor needs to remain unchanged.
Thus, the transformations may alter the behavior of the monitor arbitrarily granted the observable behavior remains the same.
\end{remark}

\section{Classical Compiler Optimizations}
\label{sec:classicOp}
In this section, we explain the adaption of classical compiler optimization techniques to the specification language \rtlola.
These techniques focuses on the expression of a stream under consideration of the pacing type.
We exemplarily introduce transformations for the \emph{Sparse Conditional Constant Propagation} and the \emph{Common Expression Elimination}.

\subsection{Sparse Conditional Constant Propagation}
\label{sec:classicOp:sccp}
Sparse Conditional Constant Propagation (SCCP) allows the programmer to write maintainable specifications without a performance penalty of constant streams.
It inlines them, pre-evaluates constant expressions, and deletes never accessed streams that includes a simple dead-code elimination.
This procedure works transitively, \ie, a stream that turns constant due to the inlining will again be subject to the same transformation.
Note that evaluating a constant expression might change the activation condition of a stream.
Thus, the transformation annotates types explicitly before changing expressions.

\subsection{Common Subexpression Elimination}
\label{sec:classicOp:cse}
The Common Subexpression Elimination (CSE) identifies subexpressions that appear multiple times and assigns the subexpressions to new streams.
These new streams might increase the required memory but save computation time by eliminating repeated computations.

In \rtlola, finding common subexpressions is simple compared to imperative programing languages for several reasons.
First, \rtlola as a declarative language is agnostic to the syntactic order in which streams are declared; the evaluation order only depends on the dependency graph.
Secondly, expression evaluations are \emph{pure}, \ie, free of side effects.
As a result, the common subexpression elimination becomes a syntactic task except that it requires access to the inferred types.
Here, two subexpressions are only considered common, if their pacing is of the same kind: periodic or event-based.
This is necessary because \rtlola strictly separates the evaluation of expressions with different pacing type kinds.

After identifying a common subexpression, the transformation creates a new stream and replaces occurrences of the expression by stream accesses.
The pacing type of the newly created stream is either the disjunction of the activation conditions of accessing streams, or the least common multiple of their evaluation frequencies.
The latter case is an over-approximation that introduces additional, irrelevant evaluations of the common subexpression.
This might decrease the performance of the monitor, so CSE is only applied if the least common multiple coincides with one of the accessing frequencies.
In this case, the transformation is always beneficial.

\section{\rtlola Specific Optimizations}
\label{sec:newOp}

This section introduces transformations around the concept of pacing types.
Since these types are specific to the specification language \rtlola, the transformations are as well.
The concepts, however, apply to similar languages as well.
We introduce the \emph{Pacing Type Refinement} and the \emph{Filter Refinement} as such transformations.

\subsection{Pacing Type Refinement}
\label{sec:newOp:pacingTypes}
In this subsection, we describe a transformation refining the pacing type of output streams.
Consider the following specification as an example.
Note, the inferred pacing types are marked gray, whereas the black ones are annotated explicitly.\\    
\begin{minipage}{\linewidth}
   	\begin{minipage}{0.47\linewidth}
   		\centering
		\begin{lstlisting}
			input alt, lat: Float64
			output check_alt $\textcolor{graynumbers}{\texttt{@\{alt\}}}$
			  := alt < $b_0$
			output check_lat $\textcolor{graynumbers}{\texttt{@\{lat\}}}$
			  := lat $\in$ [$b_1$, $b_2$]
			trigger $\textcolor{graynumbers}{\texttt{@\{alt \(\wedge\) lat\}}}$
			  $\neg$(check_alt $\wedge$ check_lat)
		\end{lstlisting}     
	\end{minipage}
	\hfill
	\begin{minipage}{0.47\linewidth}
		\centering
		\begin{lstlisting}
			input alt, lat: Float64
			output check_alt @{alt $\wedge$ lat}
			  := alt < $b_0$
			output check_lat @{alt $\wedge$ lat}
			  := lat $\in$ [$b_1$, $b_2$]
			trigger $\textcolor{graynumbers}{\texttt{@\{alt \(\wedge\) lat\}}}$
			  $\neg$(check_alt$\wedge$check_lat) 
		\end{lstlisting}      
	\end{minipage}
\end{minipage}
The specification shows a simple geofence, \ie, it checks if the altitude and latitude values are in the specified bounds.
Each expression only accesses one input stream, so the specification infers the pacing types \lstinline!@{alt}! and \lstinline!@{lat}! for the output streams.
The trigger then accesses all output stream values and notifies the user if a bound is violated.
Transitively, the trigger accesses all input streams, so its inferred pacing type is \lstinline!@{alt $\wedge$ lat}!.
With this type, the monitor evaluates the trigger iff all input streams receive a new value at the same time.
Consequently, whenever an event arrives that does not cover both input streams, the output stream computations are in vain.
This justifies refining the pacing types of the output streams to mirror the pacing type of the trigger, which is exactly what the Pacing Type Refinement transformation does.

For event-based streams, the transformation finds the most specific activation condition that does not change the observable behavior.
This goal is achieved by annotating a stream with a pacing type that is the disjunction of all pacing types accessing it.
For periodic streams, the transformation proceeds similarly.  
Here, the explicit type annotation is the slowest frequency such that each stream access is still valid, \ie, the least common multiple of each accessing frequency, similar to \cref{sec:classicOp:cse}.

Note that the pacing type transformation of a stream $s$ is only possible if all accesses to $s$ are synchronous, \ie, $(\streamany{j}, 0, \streamout{}) \in E$.
Otherwise, the transformation might change the observable behavior, as illustrated with the following example.
Consider a sliding window in a trigger condition targeting a stream $\streamout{}$.
Assume further that the transformation changes the pacing type $\outputpt{}$ from $2\hertz$ to $1\hertz$.
As a result, $\streamout{}$ produces fewer values, changing the result of the sliding window and thus the trigger as well.

The transformation resolves transitive dependencies by applying a fix-point iteration.

\subsection{Filter Refinement}
\label{sec:newOp:lazyEvaluation}
\rtlola is free of side effects and thanks to its evaluation order, it has a static program flow.
The static program flow, however, also has a drawback:  if a stream $s$ conditionally accesses a stream $s'$, $s'$ will always be evaluated before the condition is resolved.
This problem can be circumvented by integrating the condition occurring in the expression of $s$ into the filter of $s'$.

Consider the following specifications:\\  
\begin{minipage}{\linewidth}
   	\begin{minipage}{0.48\linewidth}
   		\centering
		\begin{lstlisting}
			input pilots : Float64
			input emergency : Bool
			output check_1 
			 @{emergency $\wedge$ pilots}
			  := num_pilots > 0
			  
			output check_2 
			 @{emergency $\wedge$ pilots}
			  := num_pilots == 2
			  
			trigger $\textcolor{graynumbers}{\texttt{@\{emergency \(\wedge\) pilots\}}}$
			if !emergency then check_1 else check_2
			$\phantom{a}$
		\end{lstlisting}     
	\end{minipage}
	\hfill
	\begin{minipage}{0.48\linewidth}
		\centering
		\begin{lstlisting}
			input pilots : Float64
			input emergency : Bool
			output check_1 
			 @{emergency $\wedge$ pilots}
			 { filter !emergency }
			  := pilots > 0
			output check_2
			@{emergency $\wedge$ pilots}
			 { filter emergency }
			  := num_pilots == 2
			trigger @{emergency $\wedge$ pilots}
			if !emergency then
			 check_1.hold(or: true)
			else check_2.hold(or: true)
		\end{lstlisting}      
	\end{minipage}
\end{minipage}
Both specifications check the number of pilots in the cockpit.
Depending on whether or not the plane is in emergency mode, one or two pilots are adequate.
Because of the static evaluation order, the monitor with the specification on the left always computes the values of both output streams.
However, the final trigger only uses one of the streams, depending on the \lstinline!emergency! input.
Thus, the monitor can avoid half of the output computations.
The specification on the right show how this can be achieved using Filter Refinement.
The transformation adds filters to all streams accessed in the consequence or alternative of a conditional expression.
Additionally, it replaces the synchronous lookups to these streams with asynchronous lookups and adds explicit type annotations.
The former prevents the type inference from adding the filter to the trigger as well.
The latter is necessary because the type of the trigger can no longer be inferred without the synchronous lookups.
Similar to previous transformations, Filter Refinement takes direct and transitive dependencies into account.

The algorithm for this transformation consists of four parts:
In the first step, it identifies conditional expressions.
Afterward, it constructs the filter condition for the synchronously accessed streams based on the condition following four rules.
If a stream is accesses in 
\begin{enumerate*}[label=\alph*)]
    \item the condition, it does not add any filter condition.
    \item the consequence, it adds a filter containing the if-condition.
    \item the alternative, it adds a filter containing the negation of the if-condition.
    \item a nested conditional, it builds the conjunction of the conditions.
    \item the consequence and the alternative of a nested conditional, it combines the filter conditions with a disjunction.
    \item the consequence and the alternative of a non-nested conditional, it does not add a filter. 
\end{enumerate*}
After building the filter conditions for the synchronously accessed streams, the transformation adds the filter to the stream.
If the stream already had one, the transformation builds the conjunction of both.
It then changes the affected synchronous lookups to asynchronous ones to prevent the type inference from adapting its own filter.
This process is repeated until a fix-point is reached.
Note that the transformation is only possible for synchronous lookups, otherwise the transformation alters the observable behavior.

\section{Evaluation}
\label{sec:evaluation}

We evaluate our transformations using the interpreter of the \rtlola framework~\cite{rtlolacavtoolpaper}.\footnote{\url{http://rtlola.org}}
We compare the monitor executions with enabled and disabled compiler transformations for a specification checking whether an aircraft remains within a geofence~\cite{rtlolacavindustrial}.
The traces for the evaluation consists of 10,000 randomly generated events.
Each execution was performed ten times on a $2.9\giga\hertz$ Dual-Core Intel Core i5 processor.

The geofence specification was selected due to its high practical relevance.
It checks if the monitored aircraft leaves a polygonal area, \ie, the zone for which the aircraft has a flight permission.
If the monitor raises a trigger, the vehicle has to start an emergency landing to prevent further damage.
The specification computes the approximated trajectory of the vehicle to decide whether a face of the fence was crossed.

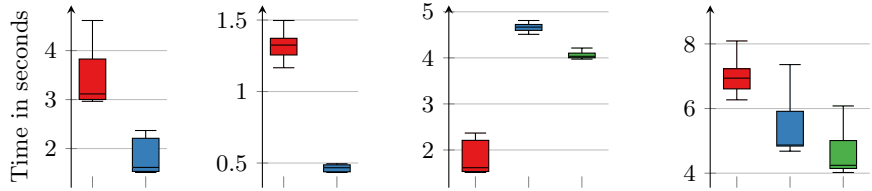
\begin{figure}[t]
	\begin{subfigure}[t]{0.215\textwidth}
		\begin{tikzpicture}
		\pgfplotstableread[col sep=comma]{dataFig1.csv}\csvdata
		\pgfplotstabletranspose\datatransposed{\csvdata}
		\pgfplotsset{%
    		width=\textwidth,
		    height=4cm,
		}
		\begin{axis}
		[
			boxplot={box extend=0.5,},
			x=0.7cm,
			enlarge x limits=0.1,
			boxplot/draw direction = y,
			x axis line style = {opacity=0},
			axis x line* = bottom,
			axis y line = left,
			enlarge y limits,
			ymajorgrids,
			xtick = {1, 2},
			xticklabel style = {align=center, font=\small, rotate=60},
			xticklabels = {},
			ylabel = {Time in seconds},
		]
			\foreach \n in {1,...,2} {
				\addplot+[boxplot, fill, draw=black] table[y index=\n] {\datatransposed};
			}
		\end{axis}
		\end{tikzpicture}
		\label{fig:eval:sccp}
	\end{subfigure}
	\begin{subfigure}[t]{0.215\textwidth}
		\begin{tikzpicture}
		\pgfplotstableread[col sep=comma]{dataFig2.csv}\csvdata
		\pgfplotstabletranspose\datatransposed{\csvdata}
		\pgfplotsset{%
    		width=\textwidth,
		    height=4cm,
		}
		\begin{axis}
		[
			boxplot={box extend=0.5,},
			x=0.7cm,
			enlarge x limits=0.1,
			boxplot/draw direction = y,
			x axis line style = {opacity=0},
			axis x line* = bottom,
			axis y line = left,
			enlarge y limits,
			ymajorgrids,
			xtick = {1, 2},
			xticklabel style = {align=center, font=\small, rotate=60},
			xticklabels = {},
		]
			\foreach \n in {1,...,2} {
				\addplot+[boxplot, fill, draw=black] table[y index=\n] {\datatransposed};
			}
		\end{axis}
	\end{tikzpicture}
	\label{fig:eval:pt}
	\end{subfigure}
	\begin{subfigure}[t]{0.275\textwidth}
		\begin{tikzpicture}
		\pgfplotstableread[col sep=comma]{dataFig3.csv}\csvdata
		\pgfplotstabletranspose\datatransposed{\csvdata} 
		\pgfplotsset{%
    		width=\textwidth,
		    height=4cm,
		}
		\begin{axis}
		[
			boxplot={box extend=0.5,},
			x=0.7cm,
			enlarge x limits=0.1,
			boxplot/draw direction = y,
			x axis line style = {opacity=0},
			axis x line* = bottom,
			axis y line = left,
			enlarge y limits,
			ymajorgrids,
			xtick = {1, 2, 3},
			xticklabel style = {align=center, font=\small, rotate=60},
			xticklabels = {},
		]
			\foreach \n in {1,...,3} {
				\addplot+[boxplot, fill, draw=black] table[y index=\n] {\datatransposed};
			}
		\end{axis}
	\end{tikzpicture}
	\label{fig:eval:lazy1}
	\end{subfigure}
	\begin{subfigure}[t]{0.275\textwidth}
		\begin{tikzpicture}
		\pgfplotstableread[col sep=comma]{dataFig4.csv}\csvdata
		\pgfplotstabletranspose\datatransposed{\csvdata} 
		\pgfplotsset{%
    		width=\textwidth,
		    height=4cm,
		}
		\begin{axis}
		[
			boxplot={box extend=0.5,},
			x=0.7cm,
			enlarge x limits=0.1,
			boxplot/draw direction = y,
			x axis line style = {opacity=0},
			axis x line* = bottom,
			axis y line = left,
			enlarge y limits,
			ymajorgrids,
			xtick = {1, 2, 3},
			xticklabel style = {align=center, font=\small, rotate=60},
			xticklabels = {},
		]
			\foreach \n in {1,...,3} {
				\addplot+[boxplot, fill, draw=black] table[y index=\n] {\datatransposed};
			}
		\end{axis}
		\end{tikzpicture}
		\label{fig:eval:lazy2}
	\end{subfigure}
\caption{
From left to right, the graphs show the impact of SCCP, Pacing Type Refinement, Filter Refinement without, and with pre-existing filters.
Red boxes are the running time before applying the respective transformation, blue after, and green by additionally applying CSE.
}
\label{fig:eval}
\end{figure}

The shape of the fence is determined statically, so the gradient and y-intercept of the faces are constants in the original specification.
We generalized the specification slightly for our case study.
This makes the specification more maintainable without forsaking performance thanks to the SCCP transformation.
In a geo-fence with five faces, the SCCP propagates and eliminates 48 constants streams.
This roughly halves the execution time of the monitor as can be seen in the first graph of \cref{fig:eval}.

In the second evaluation, we extended the specification by a third dimension, also taking the altitude of the aircraft into account.
The altitude of the aircraft is independent of the longitude and latitude, rendering computations of the output streams unobservable for events not covering all three dimensions.
Here, the Pacing Type Refinement places explicit type annotations on 32 streams in the specification with five faces.
The new trace contains a new reading for the altitude every $100\milli\second$ and for the longitude and latitude every $10\milli\second$.
The impact of the Pacing Type Refinement can be seen in the second graph of \cref{fig:eval}: the monitor for the transformed specification is roughly three times faster.

To evaluate the impact of the Filter Refinement, we adapt the specification to perform a violation check for an under-approximation of the geo-fence.  
The more costly precise geo-fence check is only performed if the under-approximation reports a violation.
This specification shows the potential impact of the Filter Refinement transformation, which adds filters to 27 output streams for a geo-fence with five faces.
The first two columns in the third graph of \cref{fig:eval} illustrate the results of the executions with a trace that is most of the time within the under-approximated fence.
Surprisingly, the specification after the transformation is about three times slower than the original specification.

The reason lies within the evaluation process of the monitor.
Filters increase the number of nodes in the dependency graph, thus triggering new evaluation steps.
In our example, this produces an overhead that is higher than the performance benefits gained by adding filters.
The last graph in \cref{fig:eval} shows the results for a specification like that for the same input trace.
Here, the transformation reduces the execution time by about 30\%.

When now also applying the CSE as well, 27 filter conditions and one if condition can be summarized in a common subexpression.
This yields another 5\% performance gain as can be seen in the last two graphs in \cref{fig:eval}.

\section{Conclusion}

Since the safety of the monitored system rests on the quality of the
monitoring specification, it is crucially important that 
specifications are easy to understand and maintain. The code
transformations presented in this paper contribute towards this
goal. By taking care of performance considerations, the
transformations help the user to focus on writing clear
specifications.

Monitoring languages are, in many ways, similar to programming
languages. It is therefore not surprising that classic compiler
optimization techniques like Sparse Conditional Constant Propagation
and Common Subexpression Elimination are also useful for monitoring.
Especially encouraging, however, is the effect of our new Pacing Type
and Filter Refinements. In our experiments, the transformations
improved the performance of the monitor as much as threefold. This
could be a starting point for a new branch of runtime verification
research that, similar to the area of compiler optimization in
programming language theory, focusses on the automatic transformation and
optimization of monitoring specifications.

In future work, our immediate next step is to integrate further common
code transformations into our framework.  We will also investigate the
interplay between the different transformations and develop heuristics
that choose the best transformations for a specific specification. A
careful understanding of the impact on the
monitoring performance is especially needed for transformations that
prolong the evaluation order, such as Common Subexpression Elimination
and Filter Refinement.

\bibliographystyle{splncs04}
\bibliography{bibliography}

\end{document}